# DETECTING DISTINCTIVE STRUCTURAL CHANGES IN ECONOMIC DATA

Joanna Dębicka[a] and Edyta Mazurek[b]

**Abstract**

This article introduces a novel method for detecting distinctive structural changes in economic data, particularly within frequency distribution tables. The approach identifies significant shifts in the distribution of a variable over time or across populations, capturing changes in category shares, enabling a deeper understanding of the underlying dynamics and trends. The method is applicable to both categorical and numerical data and is especially useful in fields such as industrial economics, demography, social science and market analysis, where comparative analysis is essential. Selected numerical examples illustrate its effectiveness in tracking market structure evolution, where shifts in firm-level market shares may signal changing competitive dynamics. The results offer interpretable insights into structural transformations in economic systems.

[a] Wroclaw University of Economics and Business, Komandorska Street 118/120, 53-345 Wrocław, Poland, e-mail: joanna.debicka@ue.wroc.pl
[b] Wroclaw University of Economics and Business, Komandorska Street 118/120, 53-345 Wrocław, Poland, e-mail: edyta.mazurek@ue.wroc.pl. Corresponding author.

## 1. Introduction

Change is a common phenomenon that affects both natural processes and social and organizational life. It is often defined broadly, with synonyms like transformation, evolution, and restructuring, reflecting its multifaceted nature (Boyne & Cole, 1998; Miller, 1982; Tushman & O'Reilly, 1996; Voldman, 2018; Zarębska, 2002). Changes can range from gradual to revolutionary, with their intensity and impact determining how they are classified. Evolutionary changes are incremental and promote stable development, while revolutionary changes are sudden and have a significant impact on the system. These can lead to innovation but may also destabilize the system if poorly managed.

Changes can also be categorized based on their effect - positive, negative, or neutral - and their depth, from superficial adjustments to fundamental transformations (Czerska, 1996; Grouard & Meston, 1997; Miller, 1982; Tushman & O'Reilly, 1996). Statistics plays a key role in analysing these changes, allowing us to identify patterns and evaluate systematic differences in structural arrays. This study focuses on identifying distinctive changes, such as shifts in the frequency of structural elements, and analysing their significance.

Traditional statistical methods (Agresti, 2017) often struggle to detect gradual, subtle changes, limiting their ability to analyse dynamic structures effectively. To address this, we propose combining these methods with additional approaches to more precisely identify and understand significant structural changes, especially those that result from dominant factors.

## 2. Assessing Structural Similarity in Frequency Tables

Considered a set of statistical units $n_X$ and $n_Y$ which form populations $X$ and $Y$ respectively. Let's assume that each statistical unit is characterized by a certain statistical feature *SF*, which takes $k$ different variants (values, classes) i.e. $sf_1, sf_2, \ldots, sf_k$. The relative numbers of statistical units in each class are denoted by the symbols $x_1, x_2, \ldots, x_k \in R$, $y_1, y_2, \ldots, y_k \in R$, where $x_1 + x_2 + \ldots + x_k = n_X$ and $y_1 + y_2 + \ldots + y_k = n_Y$. Then, vectors of the form

$$\omega(\mathbf{x}) = (\omega(x_1), \omega(x_2), \ldots, \omega(x_k)) = (x_1/n_X, x_2/n_X, \ldots, x_k/n_X), \tag{1}$$

$$\omega(\mathbf{y}) = (\omega(y_1), \omega(y_2), \ldots, \omega(y_k)) = (y_1/n_Y, y_2/n_Y, \ldots, y_k/n_Y) \tag{2}$$

defines *the simple structure of* the studied populations X and Y, respectively. In public statistics, $\omega(x_i)$ and $\omega(y_i)$ are interpreted as *indicators of structure* measuring the relative share of each class in the population and can be applied to both quantitative and qualitative features.

One of the key issues in structural analysis is to assess the similarity of two statistical communities based on their structures. This comparison can be made using distance measures between vectors $\omega(\mathbf{x})$ and $\omega(\mathbf{y})$. Thus, a measure of the similarity of structures can be an index that is a function of the distance between two structures $S = f\left(d\left(\omega(\mathbf{x}), \omega(\mathbf{y})\right)\right)$, where $f(\cdot)$ is a non-increasing function and $f(0)=1$ and $f(1)=0$ and $d\left(\omega(\mathbf{x}), \omega(\mathbf{y})\right)$ denotes a measure of the distance between structure vectors. In practice, the Bray-Curtis distance

$$d_{B-C}\left((\omega(x_1), \omega(x_2), \ldots, \omega(x_k)), (\omega(y_1), \omega(y_2), \ldots, \omega(y_k))\right) = 1 - \sum_{i=1}^{k} \min\{\omega(x_i), \omega(y_i)\} \qquad (3)$$

and distance functions of the general form $f_k\left(d\left(\omega(\mathbf{x}), \omega(\mathbf{y})\right)\right) = \left(1 - d\left(\omega(\mathbf{x}), \omega(\mathbf{y})\right)\right)^k$ where $k \in \{0.5, 1, 2\}$ are most often used to analyse the similarity of structures. Applying $k=1$ enables a clear and intuitive interpretation of the results. Based on this, *the similarity index of structures* $\omega_p$ is defined as follows:

$$\omega_p = f_1\left(d_{B-C}\left(\omega(\mathbf{x}), \omega(\mathbf{y})\right)\right) = 1 - d_{B-C}\left(\omega(\mathbf{x}), \omega(\mathbf{y})\right) = \sum_{i=1}^{k} \min\{\omega(x_i), \omega(y_i)\}. \qquad (4)$$

This index (4) is useful and flexible for analysing diverse structures. It ranges from 0 to 1, where $\omega_p = 1$ it indicates identical structures and $\omega_p = 0$ means maximum differences.

A non-parametric test based on this index, introduced by (Sokołowski, 1993), allows for assessing structural similarity (cf. Appendix A).

## 3. Identifying Distinctive Changes in Structural Tables

If only random factors influence a population, structural changes should be symmetrical, with increases and decreases balancing each other. However, if one difference stands out significantly, it likely results from a systematic factor rather than random fluctuations. The absolute value of the difference between the indicators of the structures never exceeds the magnitude of $1 - \omega_p$ i.e.

$$\max_i |\omega(x_i) - \omega(y_i)| = 1 - \omega_p. \qquad (5)$$

From equation (5), it follows that the range of possible differences between the components of the vector of simple structures is limited to the interval

$$\omega(x_i) - \omega(y_i) \in \left[\omega_p - 1, 1 - \omega_p\right]. \tag{6}$$

The closer the difference of the structure indicators is to the ends of the interval (6), the more the structures differ from each other.

Let $d_i = \omega(x_i) - \omega(y_i)$ for $i = 1, 2, \ldots, k$ be the difference between the $i$-th components of the $\omega(\mathbf{x})$ and $\omega(\mathbf{y})$ structure vectors. Furthermore, let $d_{min} = \min\{d_1, d_2, \ldots, d_n\}$ and $d_{max} = \max\{d_1, d_2, \ldots, d_n\}$. Then, we can take the sum of two intervals as the *area of distinctive absolute changes in structure*:

$$\left[\omega_p - 1, -g_p\right) \cup \left(g_p, 1 - \omega_p\right], \tag{7}$$

where $g_p = \min\{|d_{min}|, |d_{max}|\}$. Note that if $|d_{min}| = |d_{max}|$ then we do not identify any distinctive changes, we observe as profound positive as negative changes. If $|d_{min}| > |d_{max}|$ then there are classes in the structure table for which the difference between the indicators of the structure can be considered distinctive and belong to the range $\left[\omega_p - 1, -g_p\right)$. Otherwise, if $|d_{min}| < |d_{max}|$ there are distinctive differences between structure indicators belonging to the interval $\left(g_p, 1 - \omega_p\right]$.

The depth of changes depends on its nature - evolutionary or revolutionary. Distinctive changes may occur in both similar and dissimilar structures. Therefore, instead of analyzing the absolute differences $d_i$, it is reasonable to examine relative differences $r_i = d_i / g_p$. In this approach, the range of *distinctive relative changes in structure* can be defined as:

$$\left[\omega_p - 1/g_p, -1\right) \cup \left(1, 1 - \omega_p/g_p\right]. \tag{8}$$

Classes with $r_i \in [-1, 1]$ are treated as not showing distinctive changes, while those outside this range are considered structurally distinctive. The depth of such changes can be assessed using interpretive ranges of the coefficient of variation, as proposed by (Siedlecka & all, 2006). Accordingly, the interpretation of $|r_i|$ is as follows:

(1, 1.10) – statistically insignificant change,

[1.10, 1.25] – barely distinctive change,

[1.25, 1.40) – moderately distinctive change,

[1.40, 1.60] – highly distinctive change,

Above 1.60 – huge distinctive change.

## 4. Application to Market Share Data

The following section presents an empirical analysis demonstrating the practical use of the proposed method through numerical examples. The method proves effective for both similar and dissimilar structures. We apply it to cases where standard statistical tests do not reject the hypothesis of structural similarity.

Assume five companies in the market within one industry ($k \in \{A, B, C, D, E\}$) each hold equal market shares, i.e. $\omega(x_k) = 0.2$ (situation I, Table 1). During the year, hypothetical situations may occur:

- II - Company E takes 5 percentage points from D,
- III, IV – E's share is stable, while A-D shift by +/- 2-10 points (economic noise),
- V, VI - E almost exits (V) or shows the largest drop (VI), with others redistributing the share.

We begin by testing whether the structure in I is similar to those in II–VI (Table 1, columns 8-12).

Table 1. Structure arrays and structure similarity coefficients for the data from the example

| | J | I | II | III | IV | V | VI | II | III | IV | V | VI |
|---|---|---|---|---|---|---|---|---|---|---|---|---|
| (0) | (1) | (2) | (3) | (4) | (5) | (6) | (7) | (8) | (9) | (10) | (11) | (12) |
| $i$ | $sf_i$ | $\omega(x_i)$ | $\omega(y_i^{II})$ | $\omega(y_i^{III})$ | $\omega(y_i^{IV})$ | $\omega(y_i^{V})$ | $\omega(y_i^{VI})$ | \multicolumn{5}{c}{$\min\{\omega(x_i), \omega(y_i^{J})\}$} |
| 1 | A | 0.20 | 0.20 | 0.17 | 0.16 | 0.23 | 0.15 | 0.20 | 0.17 | 0.16 | 0.20 | 0.15 |
| 2 | B | 0.20 | 0.20 | 0.23 | 0.18 | 0.23 | 0.28 | 0.20 | 0.20 | 0.18 | 0.20 | 0.20 |
| 3 | C | 0.20 | 0.20 | 0.17 | 0.16 | 0.23 | 0.15 | 0.20 | 0.17 | 0.16 | 0.20 | 0.15 |
| 4 | D | 0.20 | 0.15 | 0.23 | 0.30 | 0.23 | 0.28 | 0.15 | 0.20 | 0.20 | 0.20 | 0.20 |
| 5 | E | 0.20 | 0.25 | 0.20 | 0.20 | 0.08 | 0.14 | 0.20 | 0.20 | 0.20 | 0.08 | 0.14 |
| SUM | x | 1 | 1 | 1 | 1 | 1 | 1 | 0.95 | 0.94 | 0.90 | 0.88 | 0.84 |

*Source*: own elaboration

All similarity indices (0.84-0.95) belong to the critical area, K=(0.8008,+∞), of the Sokolowski test at the 0.05 level, indicating that Market I is statistically similar to the other structures. Next, using formula (7), distinctive changes are highlighted in bold in Table 2 (columns 8-12).

Table 2. Absolute differences between structure indicators for the data in the example

| | J | I | II | III | IV | V | VI | II | III | IV | V | VI |
|---|---|---|---|---|---|---|---|---|---|---|---|---|
| (0) | (1) | (2) | (3) | (4) | (5) | (6) | (7) | (8) | (9) | (10) | (11) | (12) |
| $i$ | $sf_i$ | $\omega(x_i)$ | $\omega(y_i^{II})$ | $\omega(y_i^{III})$ | $\omega(y_i^{IV})$ | $\omega(y_i^{V})$ | $\omega(y_i^{IV})$ | | | $\omega(x_i)-\omega(y_i^J)$ | | |
| 1 | A | 0.20 | 0.20 | 0.17 | 0.16 | 0.23 | 0.15 | 0 | -0.03 | -0.04 | 0.03 | -0.05 |
| 2 | B | 0.20 | 0.20 | 0.23 | 0.18 | 0.23 | 0.28 | 0 | 0.03 | -0.02 | 0.03 | **0.08** |
| 3 | C | 0.20 | 0.20 | 0.17 | 0.16 | 0.23 | 0.15 | 0 | -0.03 | -0.04 | 0.03 | -0.05 |
| 4 | D | 0.20 | 0.15 | 0.23 | 0.30 | 0.23 | 0.28 | -0.05 | 0.03 | **0.10** | 0.03 | **0.08** |
| 5 | E | 0.20 | 0.25 | 0.20 | 0.20 | 0.08 | 0.14 | 0.05 | 0 | 0 | **-0.12** | -0.06 |
| SUM | x | 1 | 1 | 1 | 1 | 1 | 1 | 0 | 0 | 0 | 0 | 0 |
| $d_{min}$ | | | | | | | | -0.05 | -0.03 | -0.04 | -0.12 | -0.06 |
| $d_{max}$ | | | | | | | | 0.05 | 0.03 | 0.1 | 0.03 | 0.08 |
| $g_p$ | | | | | | | | 0.05 | 0.03 | 0.04 | 0.03 | 0.06 |

*Source*: own elaboration

No distinctive absolute changes were observed when comparing structure I with II and III. In comparison with IV–VI, the areas of distinctive absolute changes are as follows:

I vs IV: $[-0.1, -0.04) \cup (0.04, 0.1]$;

I vs V: $[-0.12, -0.03) \cup (0.03, 0.12]$;

I vs VI: $[-0.16, -0.06) \cup (0.06, 0.16]$.

Table 3 presents the relative differences $r_i$ between structure I and the others with distinctive changes in bold.

Table 3. Relative differences between structure indicators for the data in the example.

| | J | II | III | IV | V | VI |
|---|---|---|---|---|---|---|
| $k$ | $sf_k$ | | | $r_i$ | | |
| 1 | A | 0.00 | -1.00 | -1.00 | 1.00 | -0.83 |
| 2 | B | 0.00 | 1.00 | -0.50 | 1.00 | **1.33** |
| 3 | C | 0.00 | -1.00 | -1.00 | 1.00 | -0.83 |
| 4 | D | -1.00 | 1.00 | **2.50** | 1.00 | **1.33** |
| 5 | E | 1.00 | 0.00 | 0.00 | **-4.00** | -1.00 |
| SUM | x | 0 | 0 | 0 | 0 | 0 |

The examples presented indicate that even for similar structures, it is worth identifying distinctive changes in structure indicators. The absolute values of the relative differences differ from each other, which indicates the scale of the depth of the distinctive changes. According to the presented classification, we observe moderately distinctive differences (i.e., 1.33) and two classes where we observe huge distinctive changes (i.e., 2.5 and - 4). (cf. Table 3).

## 5. Distinctive vs. Typical and Outlying Changes

A natural question in this context is whether, from the perspective of descriptive statistics, a distinctive change is, in fact, merely an outlying or atypical observation.

Let us consider a statistical feature $D$, which takes the following values: $d_i = \omega(x_i) - \omega(y_i)$ for $i = 1, 2, ..., k$ and $\omega(x_i)$, $\omega(y_i)$ are the structure indicators of the $X$ and $Y$ features, respectively.

When analyzing the absolute differences of the structure indicators, the average value $\overline{D}$ is always equal to zero because:

$$\overline{D} = \frac{1}{k}\left(\sum_{i=1}^{k}(\omega(x_i) - \omega(y_i))\right) = \frac{1}{k}\left(\sum_{i=1}^{k}\omega(x_i) - \sum_{i=1}^{k}\omega(y_i)\right) = \frac{1}{k}(1-1) = 0. \qquad (9)$$

Let $S$ denote the standard deviation of $D$:

$$S = \frac{1}{k}\left(\sum_{i=1}^{k}(d_i - \overline{D})^2\right) = \frac{1}{k}\left(\sum_{i=1}^{k}(\omega(x_i) - \omega(y_i))^2\right). \qquad (10)$$

Typical observations lie within $(\overline{D} - S, \overline{D} + S) = (-S, +S)$, while outliers fall outside $(\overline{D} - 3S, \overline{D} + 3S) = (-3S, 3S)$, the theoretical range of the feature $D$ (Foorthuis, 2021; Hadi, 1992).

Figure 1 shows absolute differences for structures analysed in the example. Dashed lines mark typical and non-outlier ranges; red contours highlight distinctive observations.

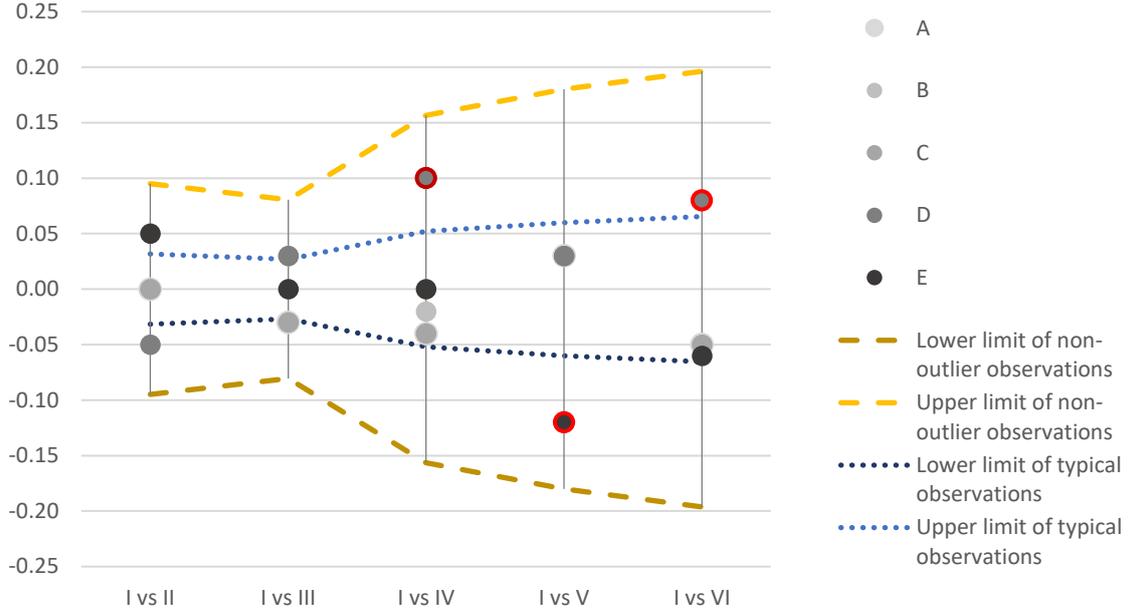

Figure 1. Descriptive statistics for the example.

*Source*: own elaboration

A distinctive change in structure indicators is not the same as an atypical observation. In cases I vs IV, I vs V, and I vs VI (marked with red contours), distinctive values were also atypical. However, not all atypical values are distinctive e.g., in I vs II, differences appear in both typical and atypical ranges without being distinctive. None of the observed values were outliers. Distinctive observations always appear opposite the remaining values along the OX axis, suggesting a link to the asymmetry of feature *D*. Let $M_3$ denote its third central moment:

$$M_3 = \frac{1}{k}\left(\sum_{i=1}^{k}(d_i - \overline{D})^3\right) = \frac{1}{k}\left(\sum_{i=1}^{k}(\omega(x_i) - \omega(y_i))^3\right), \tag{11}$$

then the asymmetry coefficient takes the following form $A = M_3/S^3$.

Table 4 presents the asymmetry indices for the situations analysed in the examples.

Table 4. Asymmetry coefficients of absolute differences of structure indices for the data of the example.

| $J$ | II | III | IV | V | VI |
|---|---|---|---|---|---|
| $A$ | 0.00 | 0.00 | 1.22 | -1.50 | 0.40 |

*Source*: own elaboration

Note that distinctive observations are those where differences in the structure indicators disrupt the symmetry of the distribution of changes. For example, a single strong increase amid several small decreases (or vice versa) leads to positive or negative asymmetry, as seen in comparisons I vs IV, I vs V, and I vs VI. These shifts signal non-random, systematic changes.

In summary, a distinctive difference is not the same as an atypical or outlier value - it reflects an asymmetry in the pattern of change.

## 6. Conclusions

The proposed method offers a refined approach to analysing structural changes by examining both absolute and relative differences in structure indicators. It enables the identification of distinctive changes, understood not as outliers or typical fluctuations, but as systematic and meaningful shifts in the distribution of a feature. Unlike traditional statistical tools, which may overlook gradual or complex modifications, this method captures both evolutionary and revolutionary changes in population structures.

The empirical example demonstrates its flexibility and practical value, especially in detecting changes that significantly affect the functioning of systems. As such, it provides analysts with a more nuanced and reliable tool for understanding the dynamics of structural transformation.

**Appendix A**

STRUCTURE SIMILARITY TEST

The structure similarity test described below was proposed in (Sokolowski 1993). The general rule of thumb for constructing and conducting a statistical test is contained in the following 6 steps.

1. Formulate the null hypothesis and the alternative hypothesis:

    $H_0$: *the similarity of the studied structures is random* (structures are dissimilar);

    $H_1$: *the similarity of the studied structures is non-random* (the structures are similar).

2. Determining the level of significance. $\alpha$

3. Choice of test statistic: the test statistic is the similarity index of structures

$$\omega_p = \sum_{i=1}^{k} \min\{\omega(x_i), \omega(y_i)\},$$

    Where *k* is the number of components of the structure.

4. Calculation of the value of the test statistic $\omega_p$ based on the sample, that is, the empirical value of the similarity index of structures, i.e. $\omega_{p,e}$.

5. Determination of the critical area: $K = (z_{\alpha,k}, \infty)$, where the values $z_{\alpha,k}$ depend directly on the significance level $\alpha$ and the number of components of the structure *k*. The exact values $z_{\alpha,k}$ are given in Table 1.

6. Decision-making:
    - when the value of the test statistic $\omega_{p,e} \notin K$ there is no basis for rejecting the hypothesis H0 that the structures are dissimilar.
    - when the value of the test statistic $\omega_{p,e} \in K$ we reject the hypothesis H0 that the structures are dissimilar in favor of the alternative hypothesis H1 that the structures are similar.

An important and useful property of the structure similarity test is that the compared structures do not have to have the same number of structure components. If there is a situation where some components are missing in the tested structures, it is then assumed that these missing components equal 0 i.e. in the case where *the i-th* relative abundance is missing in the tested variables $x_i$, it is assumed that $\omega(x_i) = 0$.

Table A1. Critical values of the similarity measure of structures

| Number of structure elements | $\alpha = 0.1$ | $\alpha = 0.05$ | $\alpha = 0.01$ | Number of structure elements | $\alpha = 0.1$ | $\alpha = 0.05$ | $\alpha = 0.01$ |
|---|---|---|---|---|---|---|---|
| 2 | 0.9362 | 0.9687 | 0.9908 | 26 | 0.5965 | 0.6195 | 0.6642 |
| 3 | 0.8377 | 0.8852 | 0.9407 | 27 | 0.5947 | 0.6173 | 0.6612 |
| 4 | 0.7897 | 0.8375 | 0.9014 | 28 | 0.5930 | 0.6152 | 0.6582 |
| 5 | 0.7550 | 0.8008 | 0.8660 | 29 | 0.5913 | 0.6132 | 0.6553 |
| 6 | 0.7280 | 0.7713 | 0.8355 | 30 | 0.5897 | 0.6112 | 0.6525 |
| 7 | 0.7064 | 0.7473 | 0.8098 | 31 | 0.5882 | 0.6093 | 0.6497 |
| 8 | 0.6889 | 0.7277 | 0.7883 | 32 | 0.5866 | 0.6074 | 0.6470 |
| 9 | 0.6747 | 0.7115 | 0.7705 | 33 | 0.5852 | 0.6056 | 0.6444 |
| 10 | 0.6629 | 0.6980 | 0.7555 | 34 | 0.5837 | 0.6038 | 0.6418 |
| 11 | 0.6532 | 0.6868 | 0.7430 | 35 | 0.5823 | 0.6021 | 0.6394 |
| 12 | 0.6450 | 0.6773 | 0.7324 | 36 | 0.5809 | 0.6005 | 0.6370 |
| 13 | 0.6381 | 0.6693 | 0.7234 | 37 | 0.5796 | 0.5989 | 0.6348 |
| 14 | 0.6322 | 0.6624 | 0.7156 | 38 | 0.5783 | 0.5974 | 0.6327 |
| 15 | 0.6271 | 0.6564 | 0.7089 | 39 | 0.5770 | 0.5960 | 0.6308 |
| 16 | 0.6227 | 0.6512 | 0.7029 | 40 | 0.5757 | 0.5946 | 0.6290 |
| 17 | 0.6188 | 0.6465 | 0.6976 | 41 | 0.5745 | 0.5933 | 0.6275 |
| 18 | 0.6153 | 0.6424 | 0.6929 | 42 | 0.5734 | 0.5921 | 0.6261 |
| 19 | 0.6123 | 0.6387 | 0.6885 | 43 | 0.5722 | 0.5910 | 0.6249 |
| 20 | 0.6095 | 0.6354 | 0.6845 | 44 | 0.5712 | 0.5899 | 0.6240 |
| 21 | 0.6069 | 0.6323 | 0.6807 | 45 | 0.5701 | 0.5890 | 0.6234 |
| 22 | 0.6045 | 0.6294 | 0.6771 | 46 | 0.5691 | 0.5882 | 0.6230 |
| 23 | 0.6023 | 0.6267 | 0.6737 | 47 | 0.5682 | 0.5874 | 0.6229 |
| 24 | 0.6003 | 0.6242 | 0.6704 | 48 | 0.5673 | 0.5868 | 0.6231 |
| 25 | 0.5983 | 0.6218 | 0.6673 | 49 | 0.5664 | 0.5863 | 0.6236 |
|  |  |  |  | 50 | 0.5657 | 0.5859 | 0.6245 |

*Source:* Sokolowski, A. (1993) *A proposal for a test of similarity of structures*. Statistical Review 40(3-4), 295-301.